\documentclass[letterpaper,prb,twocolumn,superscriptaddress,showpacs]{revtex4}
\usepackage{bm}
\usepackage{graphicx}

\begin{document}

\title{Dynamical Symmetry Enlargement in Metallic Zigzag Carbon Nanotubes}
\author{J. E. Bunder}
\affiliation{Physics Division, National Center for Theoretical
Sciences, Hsinchu 300, Taiwan}
\author{Hsiu-Hau Lin}
\affiliation{Physics Division, National Center for Theoretical
Sciences, Hsinchu 300, Taiwan} \affiliation{Department of Physics,
National Tsing-Hua University, Hsinchu 300, Taiwan}
\date{\today}
\begin{abstract}
We revisit correlation effects in doped metallic zigzag carbon nanotubes
by using both the one-loop renormalization group and
non-perturbative bosonization techniques. Note that, if a nanotube is placed
near a conducting plate, the long-range Coulomb interactions are
screened and the resulting short-range interactions can be
modelled by on-site and nearest-neighbor repulsive
interactions $U$, $V$ and $V_{\perp}$ respectively. Using both analytic and numeric means, we determine the phase diagram of the ground states. For $U/t<0.5$ ($t$ is the hopping strength),
dynamical symmetry enlargement occurs and the low-energy
excitations are described by the SO(6) Gross-Neveu model. However,
for realistic material parameters $U/t \sim \mathcal{O}(1)$, the
charge sector decouples but there remains an enlarged SO(4)
symmetry in the spin sector.
\end{abstract}
\maketitle

\section{Introduction}

Carbon nanotubes are fascinating materials with remarkable mechanical and physical properties.~\cite{review} Due to their low dimensionality, correlations and quantum fluctuations arising from electron-electron interactions cannot be ignored. It was beautifully demonstrated theoretically~\cite{Kane97} and later verified in experiments~\cite{Bockrath99,Yao99} that the unscreened Coulomb interactions drive the armchair nanotube into a Tomanaga-Luttinger liquid phase with exotic spin-charge separation. On the other hand, if the Coulomb interactions are screened so that only short-ranged interactions remain, various instabilities set in~\cite{Krotov97} and the ground state phase diagram is rather rich.

As a carbon nanotube has only a few conducting channels, it is not obvious how the long-ranged Coulomb interactions can be screened. One proposal for the reduction of long-ranged Coulomb interactions is a thick arrangement of nanotubes in either array or rope form.~\cite{Gonzalez05,Gonzalez06} Alternatively, one can make use of improving nanoscale technology and place a carbon nanotube close to a conducting plate.~\cite{Hausler02,Fogler05} The induced image charges in the conducting plate will render the Coulomb interactions $U$ short-ranged such that $U(x) \sim 1/x^3$ at long distances. The possibility of realizing short-range interactions in carbon nanotubes motivated us to revisit their correlation effects.

\begin{figure}
\centering
\includegraphics[width=7cm]{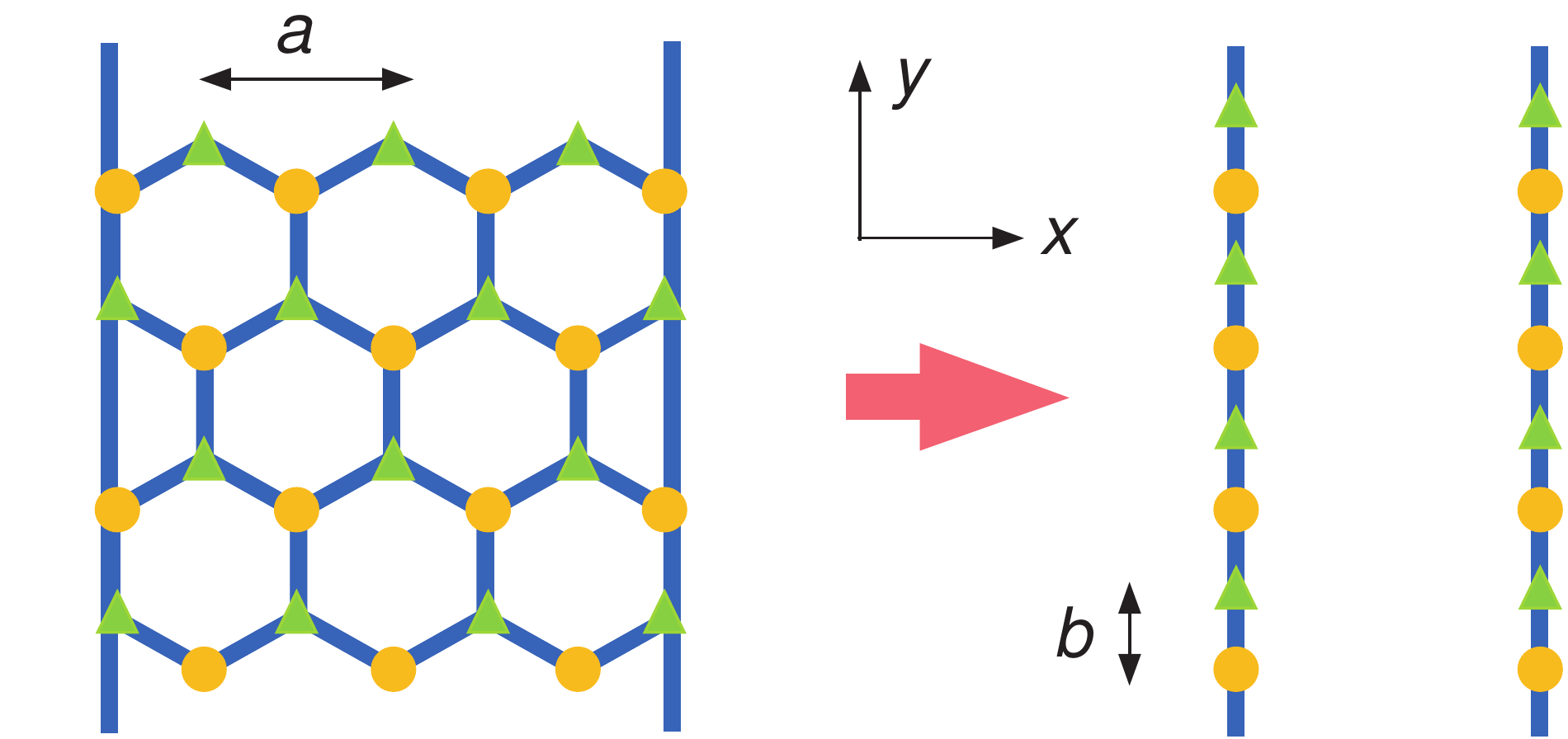}
\caption{\label{Fig1} (Color online) Zigzag carbon nanotube and
the effective two-chain model after integrating out the
higher-energy modes.}
\end{figure}

While most theoretical investigations concentrate on the armchair nanotube because of its simplicity, it is often assumed that the low-energy physics should be the same for metallic zigzag carbon nanotubes, since for both honeycomb lattices the low-energy physics originates from the Dirac cones in the band structure. We looked into this issue carefully and found that the effective lattice model of the zigzag nanotube differs from that of the armchair. Ignoring the interactions momentarily, by integrating out the gapped modes,~\cite{Lin98} the metallic zigzag nanotube is described by an effective two-chain system (shown in Fig.~\ref{Fig1}). Note that the interchain hopping is absent in this effective lattice model. As a consequence, the Fermi velocities remain degenerate when the nanotube is doped away from half filling. It is important to emphasize that the Fermi velocities in the armchair nanotube are equal to each other {\em only} at half filling and become different at finite doping. As we will explain later, the degeneracy of the Fermi velocities in the zigzag carbon nanotube is crucially important for dynamical symmetry enlargement (DSE) at finite doping. 
In addition to the subtle velocity degeneracy at finite doping, the differences between the zigzag and armchair nanotubes' effective lattice models become obvious when one tries to write the effective interactions. It turns out that, after integrating out the gapped modes, the effective interactions for the zigzag are far more complicated that those for the armchair. 

We reanalyze the correlation effects in the effective two-leg lattice model with the one-loop renormalization group (RG) and the non-perturbative bosonization techniques. Due to the spreading of the wave function, the effective interaction strength is reduced to $U_{\mathrm{eff}}, V_{\mathrm{eff}} = U/N_x, V/N_x$, where $N_x$ is the number of unit cells around the circumference and $U, V$ are the on-site and the nearest-neighbor repulsions. This reduction in the effective interaction strength is crucially important as it allows the weak-coupling RG analysis to be valid provided $U/t \lesssim 2\pi N_x$.

The one-loop RG flows determine the instabilities which may then be analyzed by non-perturbative bosonization (and refermionization) in order to identify gaps in different charge and spin sectors. From this, not only can the phase diagram be determined, but our numerical results also reveal dynamical symmetry enlargement (DSE) of low-energy excitations above the ground state.~\cite{Lin98b,Schulz98,Rice04,Chen04,Chang05} For a truly weak interaction strength $U/t<0.5$, DSE occurs and the system is described by the Gross-Neveu model with an enlarged SO(6) symmetry. The exact solution gives {\em eight-fold} kink/antikink excitations in the low-energy limit. This differs from the usual behavior of two decoupled single chains which exhibit spin-charge separation but only have four-fold degenerate spinons. It is also not the same as the fully gapped two-chain systems, where spinons and holons are glued together by long-range gauge interactions, originating from various gaps in the charge and spin sectors.

We also numerically compute the anomalous scaling of the gap/coupling ratios associated with the enlarged SO(6) symmetry. For bare couplings with approximate SO(6) symmetry, previous analytic calculations based on perturbation theory predict a universal exponent of $1/2$ for the scaling functions of the gap/coupling ratios. Surprisingly, not only do we verify the predicted exponent numerically, we also find that the exponent is robust even for large deviations in the bare couplings! Finally, for realistic material parameters $U/t \sim \mathcal{O}(1)$, the charge sector decouples and ruins
the SO(6) symmetry but there remains an enlarged SO(4) symmetry in the spin sector, indicating an unexpected degeneracy between the usual three-fold magnons and an additional neutral particle-hole excitation.

The rest of the paper is organized in the following way: In Sec II, we derive the effective lattice model for the zigzag carbon nanotube at finite doping. By carefully taking the underlying lattice structure into account, we write down the effective interactions for the on-site and nearest-neighbor repulsions. In Sec. III, we show how the field theory description can be derived from the effective lattice model by chiral-field decomposition.
Then, the bare couplings in the field-theory limit can be determined from the effective lattice model. In Sec IV, the results of the one-loop RG are shown. From the bosonization technique, the phase diagram of the ground state is obtained. We also show how DSE emerges from the numerics in weak coupling. In Sec. V, we consider the anomalous scaling behavior rising from the finite strength of interactions. Furthermore, we show that the enlarged SO(4) symmetry in the spin sector is robust as interactions increase. In the last section, we discuss the physical implications and consequences of our results.

\section{Effective Lattice Model}
\label{sec:model}

A carbon lattice may be represented by two regular triangular sublattices, offset by $\bm{d}=a(0,-1/\sqrt{3})$ and with basis vectors $\bm{a}_{\pm}=a(\pm 1/2,\sqrt{3}/2)$, where $a$ is the lattice constant. 
The Hubbard hopping Hamiltonian of such a carbon lattice is
\begin{eqnarray}
H_0 &=& \sum_{\bm{r},\alpha} \bigg\{
-t_{\perp}\: c_{1\alpha}^{\dag}(\bm{r})c_{2\alpha}(\bm{r}+\bm{d})
\nonumber\\
&&-t\: c_{1\alpha}^{\dag}(\bm{r})c_{2\alpha}(\bm{r}+\bm{a}_++\bm{d})
\nonumber\\
&&-t\: c_{1\alpha}^{\dag}(\bm{r})c_{2\alpha}(\bm{r+a_-+d})
+ h.c. \bigg\}
\end{eqnarray}
where the spatial summation goes over all lattice points $\bm{r}=n_+\mathbf{a}_++n_-\mathbf{a}_-$ with integral $n_{\pm}$. The lattice operator $c_{i\alpha}$ describes the destruction of a fermion with spin $\alpha$ in the $i$-th sublattice. The hopping amplitude along the vertical (i.e., $y$ direction) bond is $t_{\perp}$ and along the two zigzag bonds is $t$. However, in the following calculations, we only consider isotropic hopping with $t=t_{\perp}$. 

The Hamiltonian for the on-site interaction is
\begin{equation}
H_U=U \sum_{\bm{r},i} n_{i\uparrow}(\bm{r}) n_{i\downarrow}(\bm{r}),
\end{equation}
where $U$ is the on-site interaction strength and the electron density is defined by $n_{i\alpha}=c^{\dag}_{i\alpha}c_{i\alpha}$. Similarly, the nearest-neighbour interaction Hamiltonian is
\begin{eqnarray}
H_V &=& \sum_{\bm{r},\alpha,\beta} \bigg\{
V_{\perp}\: n_{1\alpha}(\bm{r}) n_{2\beta}(\bm{r}+\bm{d})
\nonumber\\
&&+V\: n_{1\alpha}(\bm{r}) n_{2\beta}(\bm{r}+\bm{a}_{+}+ \bm{d})
\nonumber\\
&&+ V\: n_{1\alpha}(\bm{r})n_{2\beta}(\bm{r}+\bm{a}_{-}+\bm{d})\bigg\}
\end{eqnarray}
where $V_{\perp}$ is the nearest-neighbor interaction strength across the vertical bond and $V$ is the nearest-neighbor interaction strength across the two zigzag bonds. In reality, we expect $V,V_{\perp}<U$. 

The Hubbard model of the zigzag carbon nanotube may be mapped onto a two-leg ladder.~\cite{Lin98} Here the $y$-axis is defined to be in the longitudinal direction of the nanotube and the $x$-axis is around the nanotube. Due to the translational invariance around the nanotube, advantage is taken of the fact that the momentum must be quantized in the $x$ direction, 
\begin{equation}
k_x=\frac{2\pi p}{aN_x},\qquad p=0,\pm 1,\ldots,\pm(N_x/2).
\label{eq:kx}
\end{equation}
The significant momenta are those which coincide with the Dirac points, i.e., the zeros of the energy spectrum. 

In the weak coupling limit $U\ll t,t_{\perp}$, it is natural to derive the effective model for the hopping first. Note that the energy spectrum for two-dimensional graphene in the tight-binding limit is
\begin{equation}
h(\mathbf{k})=2t\cos(k_xa/2)e^{ik_ya/2\sqrt{3}}
+t e^{-ik_ya/\sqrt{3}}.
\end{equation}
Therefore, the Dirac points are given by $\bm{k}=(\pm 4\pi/3a,0),(\pm 2\pi/3a,\pm 2\pi/\sqrt{3}a)$. For a metallic carbon nanotube, $N_x=3n$ and the quantized momenta cut through the Dirac points at $k_x=\pm 2\pi/3a$. For the numerics in this paper, we choose $N_x=12$ but the results hold for general metallic zigzag nanotubes.

After a partial Fourier transformation in the $x$ direction and integrating out the gapped modes, the lattice operator $c_i(\bm{r})$ is described by the fields at $k_x=\pm 2\pi/3a$,
\begin{equation}
c_i(x,y) \simeq \frac{1}{\sqrt{N_x}}
\sum_{q=\pm}d_{qi}(y)e^{iq(2\pi/3a)x}.
\label{eq:partialFT}
\end{equation}
On substitution of Eq. (\ref{eq:partialFT}) into the hopping Hamiltonian,
\begin{eqnarray}
H_0 &=& -t \sum_{y,q=\pm} [ d_{q1}^{\dag}(y)d_{q2}(y')
\nonumber\\
&&+ d_{q1}^{\dag}(y)d_{q2}(y'-2b) + h.c. ],
\label{eq:hopH}
\end{eqnarray}
where the average lattice constant is $b=a\sqrt{3}/4$ with an offset between the two sublattices $\delta=a\sqrt{3}/12$. That is to say, sublattice 1 consists of sites at $y=2mb$ and sublattice 2 at $y'=y+b-\delta$. This effective hopping Hamiltonian describes a two leg ladder with no hopping across the rungs as shown in Fig.~\ref{Fig1}. Since there is no hopping between the two legs, the Fermi velocities of the two bands remain degenerate even at finite doping. The armchair nanotube can also be captured by a similar (but different) effective two-leg model, but in this case hopping between legs is permitted and the Fermi velocities will become different when doped away from half filling.

Now let us turn to the derivation for the interacting Hamiltonians. From Eq.~(\ref{eq:partialFT}), the effective Hamiltonian from on-site repulsion is
\begin{eqnarray}
H_U &=& \frac{U}{N_x}\sum_{y,q}[n_{q1\uparrow}(y)n_{q1\downarrow}(y)
+n_{q1\uparrow}(y)n_{\bar{q}1\downarrow}(y)
\nonumber\\
&& \qquad +d^{\dag}_{q1\uparrow}(y)d_{\bar{q}1\uparrow}(y)
d^{\dag}_{\bar{q}1\downarrow}(y)d_{q1\downarrow}(y)]
\nonumber\\
&+& \frac{U}{N_x}\sum_{y,q}[n_{q2\uparrow}(y')n_{q2\downarrow}(y')
+n_{q2\uparrow}(y')n_{\bar{q}2\downarrow}(y')
\nonumber\\
&& \qquad +d^{\dag}_{q2\uparrow}(y')d_{\bar{q}2\uparrow}(y')
d^{\dag}_{\bar{q}2\downarrow}(y')d_{q2\downarrow}(y')],
\label{eq:onsiteH}
\end{eqnarray}
with $n_{qi\alpha}=d^{\dag}_{qi\alpha}d_{qi\alpha}$ and $y'=y+b-\delta$. The notation $\bar{q} = -q$ is introduced for simplicity. Note that the above Hamiltonian contains not only the density-density interaction between the two legs $q=\pm$ but also the pair-pair interaction between $(q,\bar{q})$ singlets. Similarly, we can derive the effective Hamiltonian for the nearest-neighbor repulsion,
\begin{eqnarray}
H_V &=& \frac{2V}{N_x} \sum_{y,p,q,}\sum_{\alpha,\beta}
\bigg[n_{p1\alpha}(y)n_{q2\beta}(y')
\nonumber\\
&& \hspace{-1cm} +\cos(2\pi/3) \delta_{p\bar{q}}
d_{p1\alpha}^{\dag}(y) d_{q1\alpha}(y)d^{\dag}_{q2\beta}(y') d_{p2\beta}(y')
\bigg]
\nonumber\\
&+&\frac{V_{\perp}}{N_x}\sum_{y,p,q}\sum_{\alpha,\beta}
\bigg[ n_{p1\alpha}(y)n_{q2\beta}(y'-2b)
\nonumber\\
&& \hspace{-1cm} + \delta_{p\bar{q}} d_{p1\alpha}^{\dag}(y) d_{q1\alpha}(y)
d^{\dag}_{q2\beta}(y'-2b) d_{p2\beta}(y'-2b) \bigg].
\label{eq:nnH}
\end{eqnarray}
It is important to emphasize that the factor $\cos(2\pi/3)$ comes from the underlying honeycomb lattice after a partial Fourier transformation and can not be gauged away by shifting the second sublattice by $\delta$. Furthermore, since the electrons are delocalized around the nanotube, both the effective on-site and nearest-neightbor interactions are suppressed by $1/N_x$.

The effective hopping Hamiltonian in Eq.~(\ref{eq:hopH}) and the two interaction ones in Eqs.~(\ref{eq:onsiteH}) and (\ref{eq:nnH}) now resemble an effective two-leg ladder. However, they are quite different from standard two-leg ladder Hamiltonians. The two-leg ladder Hamiltonian usually describes hopping and nearest neighbour interactions both along legs and across rungs, with the rungs positioned at right angles to the legs. However, in the zig-zag nanotube case the hopping only occurs along legs, and not across rungs. In addition, the effective interactions are far more complex than the ordinary density-density interactions. 

In particular, the inclusion of the next-nearest neighbor interaction really complicates the story since it reflects the underlying lattice structure. In fact, if one goes through the same mapping, it is straightforward to show that the effective interactions of $V$ and $V_\perp$ for the zigzag and armchair nanotubes are different. On the other hand, the effective interaction of $U$ can be shown to be equivalent by a simple (but subtle) gauge transformation.

\section{Field-Theory Limit}

To determine the ground state of the effective two-leg model, the combined usage of one-loop RG and bosonization is extremely powerful. In order to proceed, one needs to obtain a field theory description of the lattice model in the continuous limit. When considering weak interactions, a two-leg ladder is usually diagonalized so that the hopping Hamiltonian is written in terms of two decoupled bands. However, in our case the two legs are already decoupled in the hopping Hamiltonian, making any diagonalization unnecessary. In the low-energy limit, the lattice operators may be linearized about the Fermi points and expressed in terms of chiral fields. For the first sublattice, the chiral-field decomposition is rather standard,
\begin{eqnarray}
d_{q1}(y)&\approx \sqrt{b}& \bigg[\psi_{Rq}(y) e^{ik_Fy} +
\psi_{L\bar{q}}(y) e^{-ik_Fy} \bigg].
\label{eq:chiral1}
\end{eqnarray}
For the second sublattice, one may expect the finite offset $\delta$ should give rise to some phase factor in the chiral-field decomposition. However, a careful analysis leads to a somewhat surprising result,
\begin{eqnarray}
d_{q2}(y+b-\delta) &\approx& \sqrt{b} \bigg[\psi_{Rq}(y+b) e^{ik_F(y+b)}
\nonumber\\
&+&  \psi_{L\bar{q}}(y+b) e^{-ik_F(y+b)} \bigg].
\label{eq:chiral2}
\end{eqnarray}
The above result states that the chiral-field decomposition is the same as when the finite offset $\delta$ is ignored. The detailed derivation can be found in Appendix \ref{appA}.

Now we are in a position to rewrite the effective two-leg model in terms of these chiral fields. The Hamiltonian density can be separated into the kinetic and interacting parts,
$\mathcal{H}=\mathcal{H}_0 + \mathcal{H}_{I}$,
\begin{eqnarray}
\mathcal{H}_0 &=&
v\sum_{q,\alpha}[\psi^{\dag}_{Rq\alpha}i\partial_y
\psi_{Rq\alpha}-\psi^{\dag}_{Lq\alpha}i\partial_y
\psi_{Lq\alpha}],
\\
\mathcal{H}_I&=& 2\pi v \sum_{q,q'} \bigg[
c^{\rho}_{qq'}J_{Rqq'}J_{Lqq'}
-c^{\sigma}_{qq'} \bm{J}_{Rqq'} \cdot \bm{J}_{Lqq'}
\nonumber\\
&+& f^{\rho}_{qq'} J_{Rqq} J_{Lq'q'}
- f^{\sigma}_{qq'} \bm{J}_{Rqq} \cdot \bm{J}_{Lq'q'} \bigg],
\label{Hint}
\end{eqnarray}
where $v=\sqrt{3}ta/2$ is the Fermi velocity and $q, q'=1,2$ are
the chain indices in the effective model. It is worth mentioning
again that, due to the absence of the interchain hopping in the
effective two-chain system, there is only one Fermi velocity. The
electron-electron interactions are written in terms of the SU(2)
currents, 
\begin{eqnarray}
J_{Pqq'}(y) &=& \frac12 \psi^{\dag}_{Pq\alpha}(y)\psi_{Pq'\alpha}(y),
\\
\bm{J}_{Pqq'}(y) &=& \frac12 \psi^{\dag}_{Pq\alpha}(y)
\bm{\sigma}_{\alpha\beta}\psi_{Pq'\beta}(y).
\end{eqnarray}
Here $P=R,L$ denotes the chirality of the fields and $\bm{\sigma}$
is the vector of Pauli matrices.

Making use of operator product expansions for the current
products, one can derive the one-loop RG equations,
$
dg_i/dl = A_{i}^{jk} g_j g_k,
$
where $g_i$ are the couplings in Eq.~(\ref{Hint}). For the
effective two-chain system considered here, we need  {\em four}
couplings for the Cooper scattering in the charge and spin
sectors, $c^{\rho,\sigma}_{11}(=c^{\rho,\sigma}_{22})$ and
$c^{\rho,\sigma}_{12} (= c^{\rho,\sigma}_{21})$, and {\em two}
forward scattering ones $f^{\rho,\sigma}_{12} (=
f^{\rho,\sigma}_{21})$. The coupled one-loop RG equations are too
complicated to allow analytic solutions except for some special cases. The bare couplings $g_i(0)$, required for numerical integrations of RG flows, can be computed from the lattice interactions $U$, $V$ and $V_\perp$,
\begin{eqnarray}
c^{\rho}_{11} = (U+3V+3V_\perp)/t' \quad
c^{\sigma}_{11} =(U+V-V_\perp)/t'
\\
c^{\rho}_{12} = (U+3V_\perp)/t' \quad
c^{\sigma}_{12} =(U-2V-V_\perp)/t'
\\
f^{\rho}_{12} = (U+3V+3V_\perp)/t' \quad
f^{\sigma}_{12} =(U-2V-V_\perp)/t'
\end{eqnarray}
where $t' = \pi N_x v/b=2\pi N_x t$. Because of the $1/N_x$
factor, the applicability of the ``weak-coupling" regime is
extended to $U/t \lesssim 2\pi N_x$. The values of $t$ reported in
the literature~\cite{Mintmire92,Wildoer98,Odom98} range from
2.4-2.7 eV for CNTs, while $t \simeq 3$ eV is typical in
graphites. Although an accurate value of $U$ is not yet known in
nanographite systems, the value for polyacetylene, $U \simeq$ 6-10
eV,~\cite{Baeriswyl85,Jeckelmann94} might serve as a reasonable
guess. Thus, we expect that $U/t \sim 2-4$ in nanographite systems,
which satisfies the criterion for ``weak coupling".

\begin{figure}
\centering
\includegraphics[width=7cm]{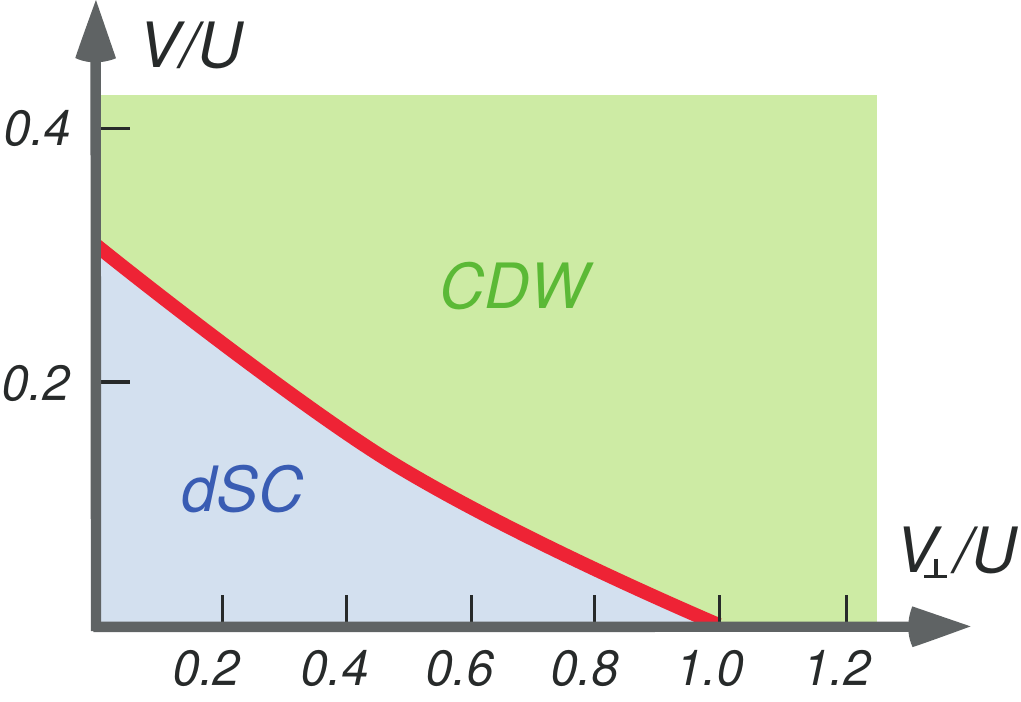}
\caption{\label{Fig2} (Color online) Phase diagram for doped
zigzag carbon nanotube with different $V_\perp/U$ and $V/U$.}
\end{figure}

\section{Dynamical Symmetry Enlargement in Weak Coupling}

To determine the phase diagram of the ground states, we first integrate the RG equations numerically up to the cutoff length scale, defined as when one of the relevant couplings reaches unity, $g_i(l_c)=1$. With $N_x =12$ and as long as $U/t \lesssim 10$, the couplings fall into two groups at $l=l_c$: the relevant couplings reach order unity and the irrelevant ones remain much smaller than unity. However, the one-loop RG alone can only determine instabilities. To truly pin down the ground state, we need to employ the non-perturbative bosonization and refermionization techniques. 
We follow the bosonization scheme described in Ref. \onlinecite{Lin98b}. The kinetic part takes the form,
$\mathcal{H}_0=\frac{v}{8\pi}\sum_\mu[(\partial_y\theta_\mu)^2
+(\partial_y\varphi_\mu)^2]$, where $\mu=\rho\pm, \sigma\pm$
denotes the total and relative charge and spin sectors
respectively, with the conjugate bosonic fields $\theta_\mu$ and
$\varphi_\mu$ describing the displacement and phase fluctuations.
The interacting part can also be bosonized after dropping the irrelevant couplings in the RG flows. For instance, the bosonized interaction Hamiltonian of the $d$-wave superconductor phase in Fig.~\ref{Fig2} is,
\begin{eqnarray}
\mathcal{H}_I &=& \frac{1}{32 \pi^2}\sum_{i=\rho+,1,2,3} K_i
\left[ (\partial_y \theta_i)^2 - (\partial_y \varphi_i)^2 \right]
\nonumber\\
&-& \sum_{i,j=1,2,3 (i \neq j)} B_{ij} \cos\theta_i\cos\theta_j,
\label{dSC}
\end{eqnarray}
where $(\theta_1, \theta_2, \theta_3) =  (\varphi_{\rho-},
\theta_{\sigma+}, \theta_{\sigma-})$. The coefficients in front of
the gradient terms are $K_{\rho+/1}= f^{\rho}_{12} \pm
c^{\rho}_{11}$, $K_{2/3} = -c^{\sigma}_{11} \mp f^{\sigma}_{12}$.
The coefficients $B_{ij}$ are symmetric with $B_{12} =
c^{\sigma}_{12}$, $B_{13} = (c^{\sigma}_{12}+c^{\rho}_{12})/2$ and
$B_{23}=-c^{\sigma}_{11}$. 

The non-zero $B_{ij}$ pin the corresponding neutral boson fields in a consistent way and
generate gaps in the $\theta_{1,2,3}$ sectors, while the total charge mode $\theta_{\rho+}$ remains gapless (described by the Luttinger liquid) and is protected by translational invariance. With the specific pinned values of $\theta_{1,2,3}$, one can show that the ground state exhibits quasi-long-range superconducting correlations with $d$-wave symmetry while all other correlation functions decay exponentially. We thus name this phase a $d$-wave superconductor (dSC). When $V$ and $V_\perp$ grow larger, the
ground state undergoes a quantum phase transition to the charge density wave (CDW) shown in Fig.~\ref{Fig2}. Since the effective Hamiltonian for the CDW is rather similar to that in Eq.~(\ref{dSC}), we will not give a detailed derivation here. Clearly, changing the interaction profile allows the system to be tuned through the dSC-to-CDW quantum phase transition.

The $1/N_x$ reduction of the effective interaction strength gives some unexpected results. For simplicity, let's start with the extremely small $U/t = 10^{-3}$ without nearest-neighbor interactions. We choose the cutoff length scale to be $|c^{\sigma}_{11}(l_c)|=1$.
Remarkably, our numerics show equality among all coefficients $K_i$ and $B_{ij}$ at the cutoff length scale. This beautiful equality among all coefficients signals an enlarged symmetry. In fact, following the refermionization scheme developed in Ref. \onlinecite{Lin98b}, one can define six Majorana fermions $\eta^{R/L}_{i}$ with $i=1,2,...,6$ to replace the three bosonic fields $\theta_{1,2,3}$. In terms of these Majorana fermions, the neutral sectors of the Hamiltonian density in Eq.~(\ref{dSC}) becomes the exactly soluble Gross-Neveu model with the enlarged SO(6) symmetry,
\begin{eqnarray}
\mathcal{H}_{N} = g_6 \sum_{a,b=1}^{6} G^{R}_{ab} G^{L}_{ab},
\end{eqnarray}
where $G^{P}_{ab} = i \eta^{P}_a \eta^{P}_b$ are the generators of the SO(6) rotations. The excitation spectrum of the SO(6) phase from the exact solution consists of two distinct features: eight kink/antikink excitations with mass $m_k$, and six fundamental fermions with mass $m_1 = \sqrt{2} m_{k}$. Above the threshold $E_c=2m_k$ is the
kink/antikink continuum.

\begin{figure}
\centering
\includegraphics[width=7.5cm]{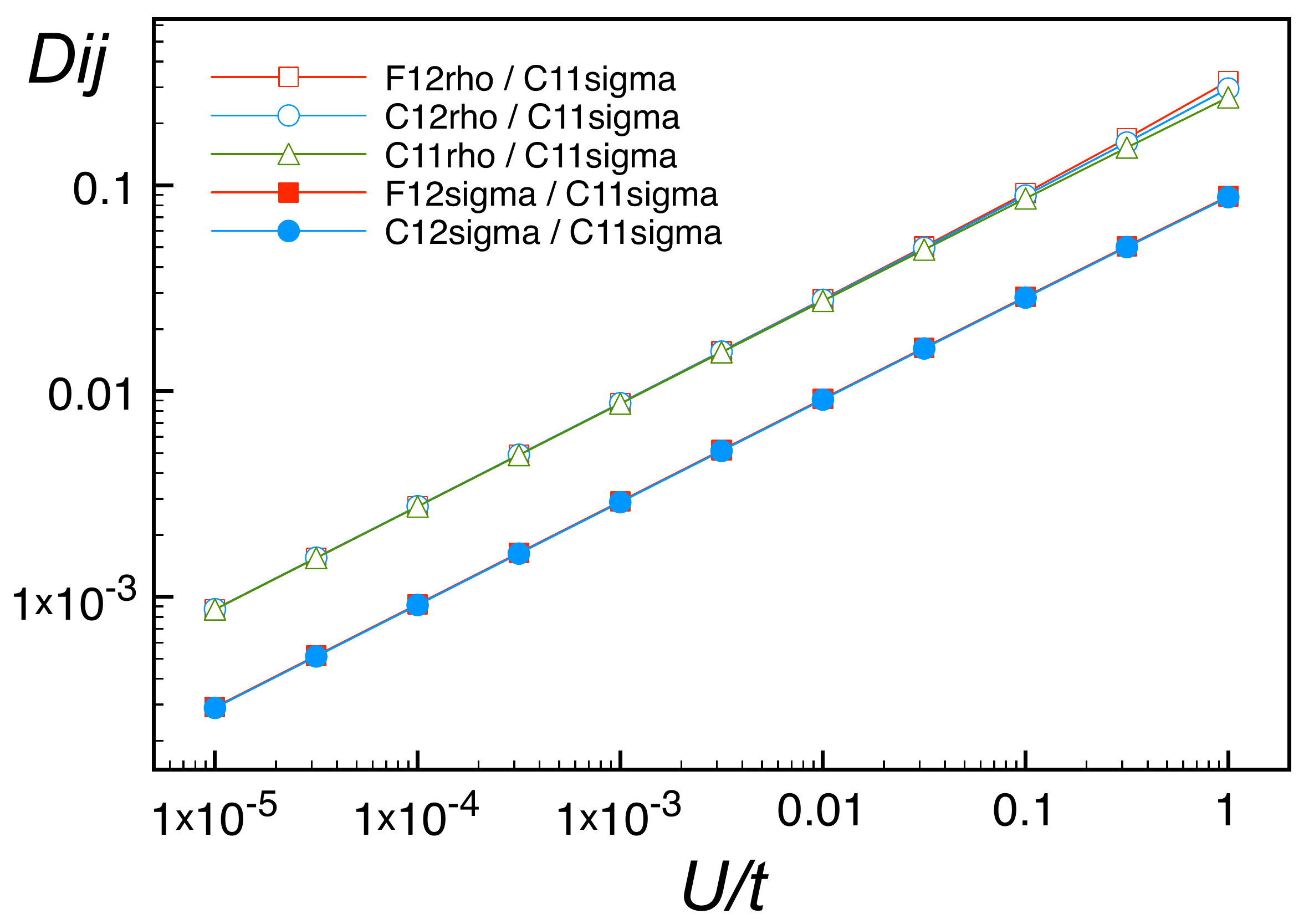}
\caption{\label{Fig3} (Color online) Log-log plot for anomalous scaling of $D_{ij}$ in weak coupling for $U/t$ ranging from $10^{-5}$ to $1$. The data collapses onto two universal straight lines with slope $1/2$, as predicted by perturbation theory.}
\end{figure}

To understand these excitations better, it is convenient to define
the conserving charges
\begin{eqnarray}
Q_{\rho-}= (Q_1-Q_2)/2, \qquad Q_{\sigma\pm} = S^z_1\pm S^z_2,
\end{eqnarray}
where $Q_{\sigma\pm}$ describes the total/relative $z$-component of the spin between two chains and $Q_{\rho-}$ describes the relative charge. The six fundamental fermions correspond to the usual three-fold degenerate magnons and three additional particle-hole bound states and carry the quantum numbers $(Q_{\rho-}, Q_{\sigma+}, Q_{\sigma-}) = (\pm 1, 0, 0), (0, \pm 1, 0), (0, 0, \pm 1)$. The kinks/antikinks carry the fractionalized quantum numbers $(\pm \frac12, \pm \frac12, \pm\frac12)$. The kinks are defined to have an even number of positive quantum numbers, while the antikinks have an odd number of positive quantum numbers. Note that, since the total charge sector decouples, all the kinks are charge neutral and can be considered to be the generalized spinons originally found in the single-chain system. However, the charge and the spin sectors are not fully decoupled here (as opposed to the single-chain system) because the kinks still carry definite
$Q_{\rho-} = \pm\frac12$. This partial spin-charge separation is reflected in the degeneracy of the spinon-like excitations. In the single-chain system there are two kink/antikink excitations, so one may expect the degeneracy to double to four-fold in a two-chain system with complete spin-charge separation. However, due to the incomplete spin charge separation the degeneracy of the kinks and antikinks in our two-chain system, and its equivalent zigzag nanotube, is eight-fold rather than four-fold.

\begin{figure}
\centering
\includegraphics[width=7.5cm]{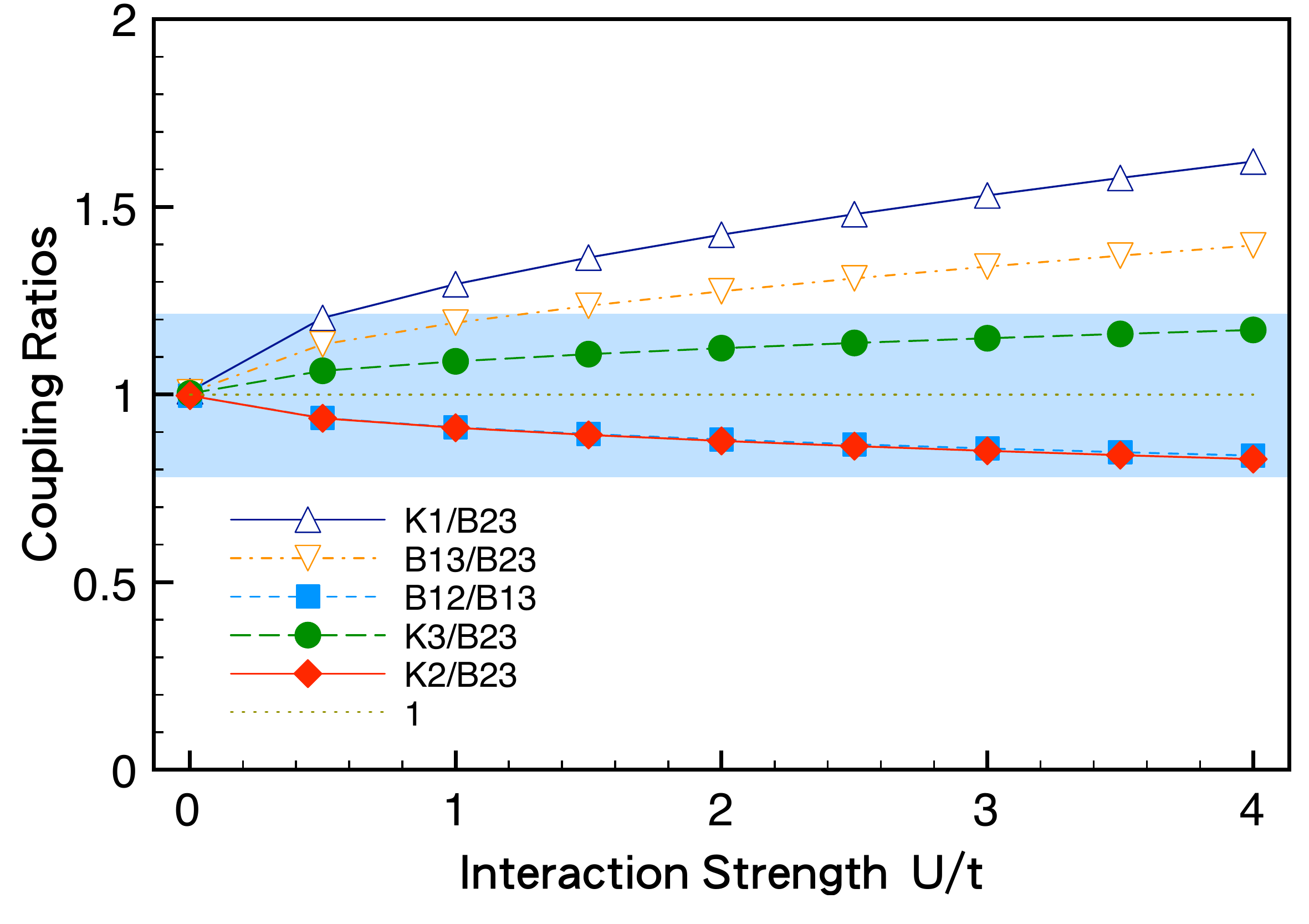}
\caption{\label{Fig4} (Color online) Coupling ratios of $K_i$ and $B_{ij}$ at the cutoff length scale for the $N_x=12$ doped zigzag carbon nanotube
with different bare interaction strength. The three filled symbols represent the relevant ratios related to the enlarged SO(4) symmetry in the spin sector.}
\end{figure}

\section{Anomalous Scaling at Finite Interaction Strength}

The unexpected enlarged SO(6) symmetry for extremely small $U/t$ is beautiful but can it survive when the interaction strength increases? Suppose $\Delta_i$ denotes the gaps for different fundamental fermions. If the SO(6) symmetry is no longer exact, the gap ratios will deviate from unity and can be calculated from the coupling ratios at the cutoff length scale.~\cite{Chang05} When DSE occurs, the equality among $K_i$ and $B_{ij}$ implies that the ratios among the couplings at the cutoff length scale are universal, $g_{i}(l_c)/g_{j}(l_c) = r_i/r_j$, where $r_i$ are order-one constants solely depending on the enlarged symmetry. Thus, it is convenient to define the deviations of the coupling ratios as a measure for the enlarged symmetry,
\begin{eqnarray}
D_{ij} = \frac{g_i(l_c)}{g_j(l_c)} - \frac{r_i}{r_j},
\end{eqnarray}
where the cutoff length scale is defined by $|c^{\sigma}_{11}(l_c)|=1$. If the bare couplings only deviate from the enlarged symmetry slightly, one can show that $D_{ij} \propto (U/2\pi t)^{1/2}$ for the SO(6) symmetry.~\cite{Chang05} Following a standard scaling argument, the gap ratios at finite interaction strength also show anomalous scaling,
\begin{eqnarray}
\frac{\Delta_i}{\Delta_j} = 1+ C_{ij} \left( \frac{U}{2\pi N_x t}\right)^{1/2}+...,
\end{eqnarray}
where $C_{ij}$ are order-one constants. The universal exponent $1/2$
is related to the enlarged SO(6) symmetry and does not depend on whether the ground state is in the dSC or CDW phase. 

However, these analytic results are not necessarily applicable to our situation. For physical short-range interactions, the bare couplings deviate from the desired SO(6) symmetry significantly and the predictions from perturbation theory may become fragile. Thus, we numerically integrate the RG flows for $U/t$ varying from $10^{-5}$ to $1$ and directly compute the coupling ratios at $l=l_c$. The numerical results are shown in Fig.~\ref{Fig3}. It is truly remarkable that the scaling form with exponent $1/2$ holds true for over five orders of the interaction strength!

Further increases in $U/t \gtrsim \mathcal{O}(1)$ will deviate from the anomalous scaling regime. Since the constants $C_{ij}$ are difficult to derive from perturbation theory, we resort to a numerical approach to determine at what interaction strength the SO(6) symmetry is no longer a good approximation. Starting from different bare interaction strengths, we compare the coefficients $K_i$ and $B_{ij}$ at the cutoff length scale, as shown in Fig.~\ref{Fig4}. For $U/t \lesssim 0.5$, all coefficients deviate by less than 20\% and the SO(6) symmetry serves as a good approximation. However, when $U/t \gtrsim \mathcal{O}(1)$, the charge sector $K_{1}$ increases rather rapidly and ruins the SO(6) symmetry. Note that this is also where the scaling form starts to break down. However, not all equalities are broken at the same time. Interestingly, even for $U/t=4$, one can read from Fig.~\ref{Fig4} that the coefficients in the spin sector remain approximately equal, $K_{2} \approx K_{3} \approx B_{23}$. A closer examination reveals that $B_{12} \approx B_{13}$ also holds.

The secret message embedded in the approximate equalities of the coefficients can be revealed by the refermionization technique. After refemionization the effective Hamiltonian now takes
the form,
\begin{eqnarray}
\mathcal{H}_{N} &=&   g_2 \sum_{AB} G^{R}_{AB} G^{L}_{AB} + g_4
\sum_{a,b} G^{R}_{ab} G^{L}_{ab}
\nonumber\\
&+& g_{24} \sum_{a,A} G^{R}_{aA} G^{L}_{aA}.
\end{eqnarray}
Here  the upper-case indices $A,B =1,2$ while the lower-case ones
$a,b = 3,4,5,6$. The SO(6) symmetry is now broken into the smaller
SO(2)$\times$SO(4) symmetry. The SO(2)$\sim$U(1) is expected
because of the conservation of the relative charge $Q_{\rho-}$.
Even though the SO(6) symmetry is destroyed in the physical regime
$U/t \sim 2-4$, the enlarged SO(4) symmetry in the spin sector is
surprising. Since the kinks/antikinks are composed of solitons in all bosonic fields, the eight-fold degeneracy still holds, although the gap ratio between the fundamental fermions and the kinks will no longer be $\sqrt{2}$ as in the SO(6) regime. Meanwhile, the six-fold degeneracy of the fundamental fermions is broken down to four-fold. An experimental investigation of these degeneracies in metallic zigzag carbon nanotubes should prove interesting.

\section{Discussions and Conclusions}

In the above sections, we study the correlation effects in doped metallic zigzag carbon nanotubes by using both the one-loop renormalization group and the
non-perturbative bosonization techniques. In particular, we are interested in the ground state properties in the presence of short-range interactions. Note that, if a nanotube is placed near a conducting plate, the long-range Coulomb interactions are
screened and the resulting short-range interactions can be
modelled by on-site and nearest-neighbor repulsive interactions $U$, $V$ and $V_{\perp}$ respectively.

By integrating out the gapped modes, the metallic zigzag nanotube can be mapped into an effective two-leg model. Due to the delocalization of electrons around the tube, the effective interactions are reduced by $1/N_x$ and thus can be treated as weak-coupling perturbations. However, since the energy of the gapped modes (which we integrated out) scales to ${\cal O}(t/N_x)$, the two-leg model will break down when $N_x$ is too large and the multibands must be included.

Using both analytic and numeric means, we determine the phase diagram of the ground states. For $U/t<0.5$ ($t$ is the hopping strength), dynamical symmetry enlargement occurs and the low-energy excitations are described by the SO(6) Gross-Neveu model. However, for realistic material parameters $U/t \sim \mathcal{O}(1)$, the charge sector decouples but there remains an enlarged SO(4) symmetry in the spin sector.

Finally, we thank the grant supports from the National Science
Council in Taiwan through NSC 94-2112-M-007-031(HHL) and NSC
93-2112-M007-005 (HHL). 

\appendix
\section{Chiral field expansion}
\label{appA}
In this appendix we derive the chiral field expansion of the fermion operator $d_{qi}$ in Eqs. (\ref{eq:chiral1}) and (\ref{eq:chiral2}). We firstly take a Fourier transform of the hopping Hamiltonian Eq. (\ref{eq:hopH}) and show that the eigenvalues are $E_{\pm}=\pm 2t|\cos kb|$ for which we can define two eigenfunctions $c_{\pm}(k)$. If $k$ is chosen to lie in the interval $[0,\pi/b]$ we can write the original fermion operators in terms of the new eigenfunctions,
\begin{eqnarray}
d_{q1}(k)&=&\frac{1}{\sqrt{2}}[c_{q+}(k)+c_{q-}(k)]\nonumber\\ 
d_{q2}(k)&=&\frac{\mbox{sign}(k-k_F)}{\sqrt{2}}e^{ik_F\delta} [c_{q+}(k)-c_{q-}(k)]
\label{eq:oddeven}
\end{eqnarray}
where the Fermi momentum is $k_F=\pi/2b$. 

If we take a spatial Fourier transform we need only retain the soft modes, which lie near the Fermi point,
\begin{eqnarray}
d_{q1}(y)&=&\frac{1}{\sqrt{2N}}\sum_p[c_{q+}(k_F+p)e^{i(k_F+p)y}
\nonumber\\
&+& c_{q+}(k_F-p)e^{i(k_F-p)y}
\nonumber\\
&+& c_{q-}(k_F+p)e^{i(k_F+p)y}
\nonumber\\
&+& c_{q-}(k_F-p)e^{i(k_F-p)y}]
\end{eqnarray}
where $0\leq p'\leq\Lambda$ for some cutoff $\Lambda$. The chiral fields are defined as
\begin{eqnarray}
\sqrt{b}\psi_{Rq}(p)&=\left\{\begin{array}{ll}
c_{q+}(k_F+p),&\qquad 0\leq p\leq\Lambda\\
c_{q-}(k_F+p),&\qquad -\Lambda\leq p\leq 0
\end{array}\right.\nonumber\\
\sqrt{b}\psi_{L\bar{q}}(p)&=\left\{\begin{array}{ll}
c_{q-}(k_F+p),&\qquad 0\leq p\leq\Lambda\\
c_{q+}(k_F+p),&\qquad -\Lambda\leq p\leq 0.
\end{array}\right.\label{eq:LR}
\end{eqnarray}
Therefore, after taking the Fourier transform of the chiral fields from $p$ to $y$ we obtain the first part of Eq. (\ref{eq:chiral1}). Note that $y=2mb$ so $e^{ik_Fy}=e^{-ik_Fy}$.

Similarly we may write $d_{q2}(y+b-\delta)$ in terms of the chiral fields. In this case we find that the finite offset $\delta$ in the spatial coordinate is cancelled by the $e^{ik_F\delta}$ which appears in the expansion of $d_{q2}(k)$ in Eq. (\ref{eq:oddeven}). Therefore, the lattice operator in terms of $c_{q\pm}(k_F\pm p)$ is
\begin{eqnarray}
d_{q2}(y+b+\delta) &=&\frac{1}{\sqrt{2N}} \sum_p 
[c_{q+}(k_F+p)e^{i(k_F+p)(y+b)}
\nonumber\\
&-& c_{q+}(k_F-p)e^{i(k_F-p)(y+b)}
\nonumber\\
&-& c_{q-}(k_F+p)e^{i(k_F+p)(y+b)}
\nonumber\\
&+& c_{q-}(k_F-p)e^{i(k_F-p)(y+b)} ]
\end{eqnarray}
once all but the soft modes have been discarded. Substituting Eq. (\ref{eq:LR}) and taking the Fourier transform gives the second part of Eq. (\ref{eq:chiral2}).

\end{document}